\documentclass{aa}
\usepackage{graphicx}
\usepackage{txfonts}
\begin{document}
   \title{HD 65949: The  Highest Known Mercury
   Excess Of Any CP Star?}

   \author{Charles R. Cowley
          \inst{1}
          \and
          S. Hubrig\inst{2}\fnmsep\thanks{
          Based on ESO programe  076.D-0172(A).}
          \and
          G. F. Gonz\'{a}lez\inst{3}
          \and
          N. Nu\~{n}ez\inst{4}
          }

   \offprints{C. Cowley}

   \institute{Department of Astronomy, University of Michigan,
              Ann Arbor, MI 48109-1090, USA\\
              \email{cowley@umich.edu}
         \and
             European Southern Observatory,
             Casilla 19001, Santiago 19, Chile\\
             \email{shubrig@eso.org}
         \and
             Complejo Astron\'omico El Leoncito, Casilla 467,
             5400 San Juan, Argentina\\
             \email{fgonzalez@casleo.gov.ar}
          \and
             Departamento de Geof\'{\i}sica y Astronom\'{\i}a, Facultad de Ciencias Exactas
             F\'{\i}sicas y Naturales,
             Universidad Nacional de San Juan, Argentina
             }

   \date{Received ; accepted }

\abstract{ESO spectra of HD\,65949 show it to be
unlike any of the
well-known types within its temperature range
$\approx$ 13600K.  It is
neither a silicon, nor a mercury-manganese star, though
it has a huge Hg II line at $\lambda$3984.  We
estimate $\log({\rm Hg/H}) + 12.0 \approx 7.4$.  This
is higher than any published stellar mercury abundance.
HD\,65949 is a member of a nearby open cluster, NGC\,2516, which
is only slightly
older than the Pleiades, and has been of recent interest
because of its numerous X-ray emission stars, including
HD\,65949 itself, or a close companion. A
longitudinal magnetic field
of the order of $-290$ Gauss at the
4.7~$\sigma$ level was very recently diagnosed from accurate circular
spectropolarimetric observations with FORS 1 at the VLT.

The spectral
lines are sharp, allowing a thorough identification study.
Second spectra of Ti, Cr, and Fe are rich.  Mn II is well
identified but not unusually strong.  Numerous lines of
S II  and P II are found, but not Ga II.  The resonance
lines of Sr II are strong.  While many Y II lines are
identified, and Nb II is very likely present, {\it no}
Zr II lines were found.  Xe II is well identified.  Strong
absorptions from the third spectra of the lanthanides  Pr,
Nd, and Ho are present, but lines from the second spectra of
lanthanides are extremely weak or absent.
Among lines from the heavier
elements, those of Pt II are clearly present, and the heaviest
isotope, $^{198}$Pt, is indicated.  The uncommon spectrum of Re II is
certain, while Os II and Te II are highly probable.  Several
of the noted anomalies are unusual for a star
as hot as HD 65949.
            }
\titlerunning{A New Class of CP Star?}
\authorrunning{Cowley et al.}

\maketitle
%

\section{Introduction}
\label{se:intro}

HD 65949 may have the highest mercury abundance of any known
star.  The unusual strength of Hg II $\lambda$3984 was already
noted by Abt and Morgan (1969) in their study of the open cluster
NGC 2516.  They did not call it a manganese
star.

The age of the cluster was estimated as log\,$t$ = 8.2 $\pm$ 0.1
(Sung, et al. 2002).
Several studies indicate an X-ray source closely
coincident with the position of HD 65949 (cf. Damiani, et al
2003, Table 4, Source No. 44).

Chemically peculiar (CP) stars in the temperature range corresponding
to mid-B to A0 are usually either silicon or mercury-manganese
stars.  The former belong to the magnetic sequence,
designated CP2 by Preston (1974).



Line identifications for HD 65949 were made with the help
of predicted line
strengths from a model atmosphere with $T_{\rm eff} = 13600K$
and $\log(g) = 4.0$, derived from Str\"{o}mgren colors
and H$\beta$ photometry (cf. Moon, 1984).
Artificially high abundances of ``metals'' were assumed
in order to avoid missing lines.
Central intensities greater than 0.001 were derived for
all lines ($\lambda\lambda$3500-9400)
in VALD (Kupka, et al. 2000).
The technique was presented by Cowley (1995) and
is similar to one available to users of VALD.

A few abundances are derived below.  These are based
on a provisional or temporary model with
the same $T_{\rm eff}$ and log($g$) from
Str\"{o}mgren photometry, but with
abundances enhanced by a factor of 3 for atomic
numbers (Z) from 3 to 35, a factor of 10 for Z from
36 to 72, and 100 for Z above 73.  Following Dworetsky
(2005), we assumed a depleted helium abundance,
He/H = 0.0085.

We have placed a detailed line identification list
with additional references online:
http://www.astro.lsa.umich.edu/~cowley/hd65949.
We also provide links to two tables, highlighting
the Pt II and Re II spectra.


\section{Observations}

Spectra of HD 65949 were obtained on 4 consecutive nights,
from 20 to 23 October 2005,
with the FEROS echelle spectrograph on the 2.2m telescope
at La Silla.
Basic steps of spectroscopic reduction (bias subtraction,
division by a normalized flat field, extraction of a 1D spectrum,
continuum normalization and wavelength calibration)
were based on IRAF.

The wavelength coverage is from 3530 to 9220~\AA,
and has a nominal resolving power of 48 000, with a S/N
ratio of ca. 200-250.  The spectra were mildly Fourier
filtered before the wavelengths were measured at Michigan.

A single longitudinal magnetic field measurement in
HD\,65949 was recently carried out
by Hubrig et al. (2006) using FORS\,1 at the VLT
in its spectropolarimetric mode.
A weak magnetic field of the order $-290$ Gauss at the
4.7~$\sigma$ level was detected.

\section{Manganese, Mercury, and Platinum}

\begin{figure}
   \centering
   \includegraphics[width=0.30\textwidth,angle=270]{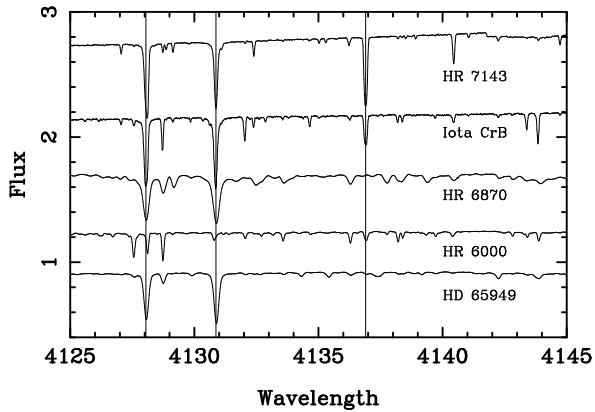}
      \caption{Thin vertical lines are at the wavelengths of
          the strong Si II doublet,
          $\lambda\lambda$4128 and 4130, and Mn II $\lambda$4137,
          which appear as a ``triplet'' in the upper two
          classical HgMn stars.  The lower three stars are
          neither HgMn nor silicon (CP2) stars
          (cf. Section~\ref{sec:outlier}).
          Note the weakness
          of the Si II lines in HR 6000, and the Mn II line in
          HR 6870 and HR 65949.  The absorption features are
          not precisely
          bisected because of significant blending.
          }
         \label{fig:one}
   \end{figure}

At higher dispersion,
HgMn stars may be recognized by the presence of
Mn II $\lambda$4137, which forms an easily recognizable triplet
with the nearby Si II doublet.  This triplet is illustrated
in Fig.~\ref{fig:one}.  It is readily seen for the star labeled
NGC 2516B
on Plate 17 of the atlas by Morgan, Abt, and Tapscott (1978),
based on 39 \AA/mm spectrograms.
The Mn II line does not appear in the Multiplet
Tables (Moore, 1945).

%
%

One sees immediately from Fig. 1 that HD 65949 cannot be classified
as a manganese star.  The strong Mn II line $\lambda$4137 is at
best marginally present.  Mn II, $\lambda$4137 is somewhat weaker
in the cooler HgMn star $\iota$ CrB than in the hotter HR 7143.
Generally, though not always,
the Mn II spectrum strengthens toward the hotter HgMn types,
reaching a maximum near 14 000K.   Thus HD 65949 is near the
temperature where Mn II would be expected to reach a maximum.

\begin{figure}
   \centering
   \includegraphics[width=0.30\textwidth,angle=270]{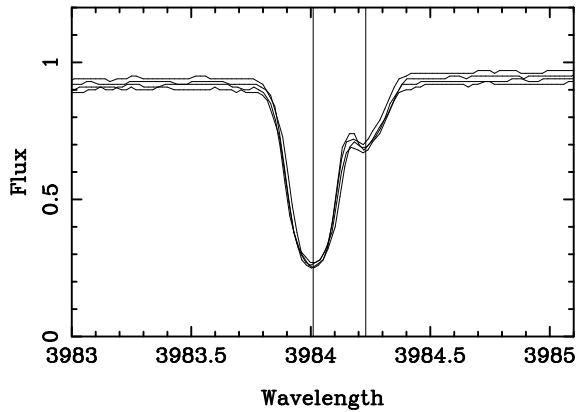}
      \caption{Profiles of the $\lambda$3984-region obtained
          on 4 sequential nights.  They have been shifted to
          bring the mean wavelength of the strong component
          to 3984.01~\AA.  Adopted continuua from the
          original reductions have not been modified.  The feature
          at 3984.23~\AA, is primarily due to Re II (see text).
          }
         \label{fig:two}
   \end{figure}

The Hg II line  $\lambda$3984, however,
has a huge equivalent width (cf. Fig.~\ref{fig:two}).
With one possible exception,
it is larger than any of the values
listed for the 43 stars included in the studies of White,
et al. (1976), or Woolf and Lambert (1999).  The possible
exception is the value Woolf and Lambert give, 177~m\AA\_
for HR 2844.  They note that
White, et al. give a smaller value: $W_\lambda = 144$~m\AA.
The star has
substantial rotational broadening,
$v\cdot\sin(i) \approx 30$, making an accurate measurement
of the equivalent width difficult.
%

We estimate
$W_\lambda(3984) \approx$ 175 m\AA, using the wing of
H$\epsilon$ as the (local) continuum, and
excluding
the nearby absorption in the red wing
(cf. Fig.~\ref{fig:two}).
This line has a distinct minimum, at
3984.23$~\pm 0.02$ \AA , too long for it to be
$^{204}$Hg II, $\lambda$3984.07
(We adopt Hg isotopic wavelengths from Woolf and Lambert).
Our measured wavelength
for the minimum of the strong feature is
$3984.01~\pm 0.01$~\AA, indicating that $^{204}$Hg must be
a major contributor to the strong absorption--it is not the
weaker absorption on the red wing.  A calculated profile
based on (1/6)$^{200}$Hg, (1/2)$^{202}$Hg, (1/3)$^{204}$Hg,
and ignoring other isotopic contributions fit the profile
reasonably well.

HD 65949 is known to be a spectroscopy binary (cf. Gieseking
and Karimie 1982), but we have been unable to find absorption
lines that are definitely due to the secondary.
The line at $\lambda$3984.23 is primarily due to Re II,
though there may be some contribution from Cr II.  We
discuss the Re II identification in a separate section.

\begin{figure}
   \centering
   \includegraphics[width=0.30\textwidth,angle=270]{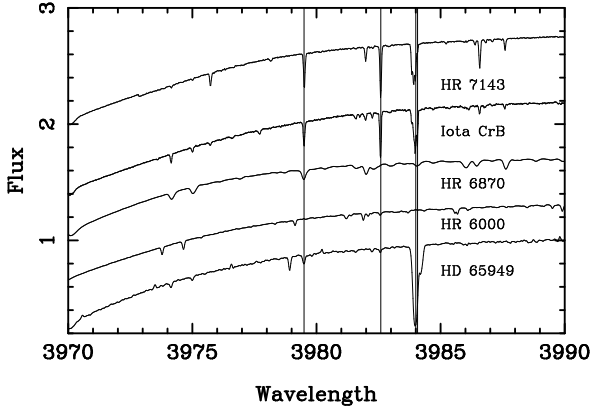}
      \caption{Thin vertical lines are at the wavelengths of
          $^{204}$Hg II, $\lambda\lambda$3984.07;
          $^{200}$Hg II, 3983.91; 3982.59 (Y II-6), and
          3879.50 (Cr II-183).
          }
         \label{fig:hgreg}
   \end{figure}

Fig.~\ref{fig:hgreg} shows the wider
wavelength region of $\lambda$3984 for the 5 stars of
Fig.~\ref{fig:one}.
%
%

The red 7s-7p
Hg II lines,  $\lambda\lambda$6149 and 7944 are
also present.
Additionally, all three expected
Hg I lines, $\lambda\lambda$4046, 4358, and 5460 are
present, within 0.02~\AA\ of their
laboratory positions (cf. Table~\ref{tab:pab}).
We measured $\lambda^*$4046.54 for the
first of these lines.  It is blended with Pt II, whose wavelength
could be anywhere from 4046.44 to 4046.52~\AA\ depending on
the isotopic composition (Wahlgren, et al. 2000).


   \begin{table}

     \caption{Provisional Abundances from Hg I and II}
         \label{tab:pab}
   \begin{tabular}{l c c c c} \hline\hline
Spectrum\rule[-0.03in]{0.0in}{0.2in}
&$\lambda$(VALD)&$\lambda$(HD 65949)&$W_\lambda$(m\AA)&Log(Hg/H)\\    \hline
Hg I & 4358.32 &58.35&22&-4.57  \\
Hg I & 5460.73 &60.73&29&-4.48  \\
Hg II& 6149.47 &49.46&41&-4.66  \\
Hg II& 7944.55 &44.57&23&-4.81 \\  \hline
\end{tabular}
\end{table}

We derive an abundance estimate for mercury based on our
temporary model, using
the lines shown in Table~\ref{tab:pab}.  The values are
all within a factor of two of one another.  A straight
mean yields log(Hg/H) + 12.0 = 7.4.  Our rough calculation
for $\lambda$3984, with a three-isotope mix,
achieves a fit for an abundance between 7.1 and 7.4.

A value log(El/H) = 7.4 is close to the
solar {\it iron} abundance, and
higher by 0.75 dex than in any of the 42 stars studied by Woolf and
Lambert (1999).  Dolk, et al. (2003) give mercury abundances
for 31 HgMn stars including 14 in common with Woolf and Lambert.
Their highest abundance is 7.0, for AV Scl (HR 89 $=$ HD 1909).

In addition to mercury, we
also identify 15 lines wholly or partially as Pt II.  The stellar
wavelengths are in better agreement with laboratory wavelengths
if we apply double the isotope shifts given by Engleman (1989)
for $^{196}$Pt vs. $^{194}$Pt to get wavelengths for the heaviest
stable isotope, $^{198}$Pt (cf. Dworetsky 1993).
Table~\ref{tab:pt} gives data for 10 of the 15 lines
where we think Pt II dominates the stellar feature.
The first two columns give the measured stellar wavelength,
and Shenstone's (1938) intensity.
The entry, $\Delta_S$, in Column 3 is
the stellar wavelength minus Shenstone's wavelength.

Column 4 gives $\Delta_{198}$, the stellar wavelength
minus that for $^{198}$Pt II, using Engleman's isotope
shifts.  Both differences are in hundredths of an angstrom.
The equivalent
width is given in Colunmn 5, followed by the corresponding
abundance based on oscillator strengths from Dworetsky,
Storey, and Jacobs (1984).  Possible blends are listed
in the final column.  The platinum abundance excess
is $\approx$ 4.6  dex.


   \begin{table}[h]
     \caption{Pt II Lines with Minimal Blending}
         \label{tab:pt}
   \begin{tabular}{l c c c c l l} \hline\hline
$\lambda^*$\rule[-0.03in]{0.0in}{0.2in}
       &${\rm I_S}$
            &($\Delta_S)$
                &($\Delta_{198}$)
                     &$W_\lambda$
                                  &log
                                        &  Blend \\
\AA    &    &\AA$\times$100
                &\AA$\times$100
                     &m\AA        &(Pt/$\Sigma$) &   \\  \hline
3766.46& 10 & 6 & -2 &10$^{\rm B}$&     &  \\
3768.45& 10 & 6 & -1 &10$^{\rm B}$&     &  \\
4023.85& 3  & 4 &  1 &17          &-5.74&    \\
4034.21& 5  & 4 & -1 &23          &-5.82& .23 Fe II      \\
4061.71& 10 & 5 &  3 &37          &-5.58& .78 Fe II      \\
4105.50& 0  & 5 &  1 &6$^{\rm B}$ &     &  \\
4148.33& 5  & 3 &  2 &17          &-5.79&                \\
4223.76& 3  & 7 &  5 &12          &     &  \\
4288.46& 5  & 6 &  1 &25          &     &.56 Fe II?  \\
4514.18& 10 & 1 & -1 &52          &-5.69&   \\  \hline
\end{tabular}

\footnotesize{$^{\rm B}$Line is in a Balmer wing}
\end{table}


Note that
Pt II is commonly identified in the {\it cooler} HgMn stars,
and that a dominant heavy isotope (Pt, or Hg) is also
more typical of the cooler stars.

We find other similarities to HgMn stars.  Both S II and P II are
well identified.  These ionic spectra appear sometimes individually,
and sometimes together in HgMn stars.
P II is not commonly identified in the
magnetic sequence of CP stars.
We find no evidence for Ga II, which
often is found in stars with strong P II.

\section{The Re II Spectrum}

We found highly significant coincidences with
{\it classified} Re II lines having
intensities of 20 or greater (Meggers, Catal\'{a}n,
and Sales 1958).   After noting
that the Meggers, et al. paper included a line at 3984.28~\AA,
with intensity 10,
we tested significances with an {\it independent} list of 19
lines, all with intensity 10.  The results were
again highly significant. Moreover,
the additional lines with intensities
10 and 15 filled in a number of {\it previously unidentified}
wavelengths.

We were able to identify additional weak features with
rhenium, probably Re II, with the help of Meggers's (1952)
description of the arc and spark spectrum of the element.
Thus far, we have identified nearly 60 lines as due wholly
or partially to rhenium.  A detailed listing may be
found in the online material.

The Re II spectrum has been identified in CP stars, but with
difficulty, and some uncertainty.  Palmeri, et al. (2005) give
details, as well as a few oscillator strengths.  The strongest
Re II line for which they give
an oscillator strength is $\lambda$3742.28, log($gf$) = $-$1.7.
The equivalent width of $\lambda^*$3742.30 is 74~m\AA.  If the
line is unblended, a provisional abundance estimate yields
$\rm \log(Re/H) + 12 \approx 6.2.$,
(using $\rm \xi_t = 2k s^{-1}$ to
simulate hyperfine structure).  This is somewhat
lower than the mercury abundance ($\approx$ 7.4).
The transition probability
is given only to 1 figure; the overall calculation is
quite rough.

\section{Relation to the silicon stars}

HD 65949 has a weak magnetic field, and therefore some
similarity to the CP2 stars.  Its temperature puts it in the
range of the silicon stars, but the silicon lines themselves
are not particularly strong (cf. Fig. 1).

Among the known silicon stars
with temperatures near 14000K, we might compare it with
HD 200311, a mild silicon star studied by Adelman (1974).
HD 200311 has sharp spectral lines, and shows Hg II
$\lambda$3984, for which we
measured equivalent widths
of 69 and 90~m\AA~\ on two
analog tracings kindly made available by Dr. Adelman.

The $\lambda$3984 line of Hg II is identified in a number
of silicon stars (cf. Albacete-Colombo, et al. 2002).

HD 200311 has too strong Co II lines for it to be considered
a typical CP2 star, but in any case its myriad Ce II lines
set it clearly apart from HD 65949, which has no securely
identified lines of the second spectrum of any lanthanide.

Another hot silicon star with published equivalent width
measurements is HR 1728 (HD 34452, Tomley, Wallerstein, and
Wolff 1970).  The spectrum has substantial rotational
broadening which may have obscured the Hg II or most
lines, but Eu II $\lambda\lambda$3930 and 4205 have
measured equivalent widths of 80 and 120~m\AA.  The Si II
lines themselves, are {\it much} stronger than in
HD 65949.

\section{The outliers: HR 6000 and HR 6870}
\label{sec:outlier}

The silicon-weak star HR 6000 is currently subject to a detailed
study by Castelli (2006).  She has used a temperature of
12850K, close to that appropriate for HD 65949.  However,
Fig. 1 shows the most extraordinary
weakness of the typically strong Si II lines
$\lambda\lambda$4128 and 4130.  Given the ubiquity and common
strength of these lines for stars in this temperature range,
we conclude that HR 6000 is significantly different from
HD 65949 and its congeners at this temperature.

Another unusual star with a comparable temperature is
HD 6870 (HD 168733), noted by Bidelman  and Aller (1963).
Little (1974) and
Muthsam and Cowley (1984) determined abundances.  These workers
noted unusually strong Cl II, and Ga II, and Ti II and III.
The iron abundance is quite high.  Muthsam and Cowley
noted the presence of the third spectrum of Nd.  A new
study in progress,
of UVES spectra extending to 3048\AA, shows that
Dy III, Sm III, and most unusually, Ce III are clearly
present.

While HD 64949 has some features in common with HR 6870,
the spectra are quite different.

\section{Marginal identifications}
\label{sec:marg}
We analyzed 2264 wavelength measurements by the method
of wavelength coincidence statistics.  The survey employed
416 lists of atomic and ionic lines with various strengths.
The number of spurious marginal coincidences to be expected
may be determined by a Monte Carlo method.  We expect
between 8 and 12 marginally significant results, by chance.
We obtained 25 marginal significances (0.01 to 0.05) for
the following spectra: He I, N II, Ne I (2 lists), Sc II
(2 lists), Ti II, V I, Co I, Ni II (2 lists), Cu II,
Kr II (2 lists), Sb II, Nd III (2 lists), Tb III, Dy III,
Au II (3 lists), Hg II, Pu II, and Cm II.  Several of
these species are surely present (e.g. He I, and Nd III).

The following spectra showed highly significant coincidences
($<$ 0.001): Re II (2 lists), Os II, Te II (2 lists), Ru II,
and Nb II (2 lists).  These spectra are rarely identified
in stars and are of great interest.

\section{Discussion}

HD 65949 is one of a number of CP stars in the cluster
NGC 2516, which is slightly older than the Plieades.  It
is unclear how this fact might be related to the
peculiarities discussed here.


Although HD 65949 probably has the highest mercury
abundance of any known CP star,
it is not a mercury-{\it manganese} star.
Typically, HgMn stars with similar temperature have
the greatest overabundances of manganese.  Whatever
mechanisms operate to produce the overabundances of
the heaviest elements, have failed to produce the
usual anomalies near the iron peak.

Clues to the unusual nature of HD 65949 may lie hidden
among the abundances of trace and marginally identified
species for which superior observational material is
required.

\begin{acknowledgements}

This research has made use of the SIMBAD database,
operated at CDS, Strasbourg, France.  We acknowledge
use of the Vienna Atomic Database (VALD).

We thank Dr. S. J. Adelman for tracings of HD 200311,
and B. Smalley for a copy of code to determine stellar
parameters.  Dr. Luca Sbordone kindly helped CRC port a
Linux Atlas 9 to a pc running Digital
Visual Fortran.  CRC thanks Dr. Jon Miller for a helpful
discussion of the X-ray sources in NGC 2516.

\end{acknowledgements}

\end{document}